\documentclass[12pt,journal]{IEEEtran} 
\ifCLASSINFOpdf
\else
\fi

\usepackage{color}
\usepackage{amssymb}
\usepackage{amsthm}
\usepackage[cmex10]{amsmath}
\DeclareMathOperator{\E}{\mathbb{E}}

\newcommand {\Define} {\stackrel {\Delta} {=}  }

\newcommand {\pu} {p_{\text{u}}}

\usepackage{bm}
\usepackage{mathtools}
\usepackage{fixltx2e}
\usepackage{epsfig,makeidx,color}
\usepackage{graphicx,dblfloatfix}
\usepackage{amsbsy}
\usepackage{amssymb}
\usepackage{euscript}
\usepackage{lipsum}
\usepackage{cite}
\usepackage{chngcntr}
\usepackage{lineno}
\usepackage[T1]{fontenc}
\usepackage{microtype}
\usepackage{wasysym}
\usepackage{footnote}
\usepackage[left=01 in,top=0.5 in,right= 1 in,bottom=0.5 in]{geometry}        

\newtheorem{remark}{\it Remark}

\usepackage[english]{babel}
    \usepackage[font=small,labelsep=space]{caption}
\usepackage{subcaption}   

\makeatletter
\def\citenoauxwrite#1{\begingroup
\@fileswfalse
\cite{#1}\relax
\endgroup}
\makeatother

\begin{document}

\title{Information Rate Performance of Massive MU-MIMO Uplink with Constant Envelope Pilot-based Frequency Synchronization}
%
%
%

\author{Sudarshan~Mukherjee~ and 
        ~Saif Khan~Mohammed
\thanks{The authors are with the Department of Electrical Engineering, Indian Institute of Technology (I.I.T.) Delhi, India. Saif Khan Mohammed is also associated with Bharti School of Telecommunication Technology and Management (BSTTM), I.I.T. Delhi. Email: saifkmohammed@gmail.com. This work is supported by EMR funding from the Science and Engineering
Research Board (SERB), Department of Science and Technology (DST),
Government of India.}
\thanks{This paper is a substantial extension to the work submitted to IEEE Globecom 2016 \protect\citenoauxwrite{gcom2016}.} 
}

\onecolumn
\maketitle

\vspace{-2.5 cm}

\begin{abstract}
In this paper, we consider a constant envelope (CE) pilot-based low-complexity technique for frequency synchronization in multi-user massive MIMO systems. Study of the complexity-performance trade-off shows that this CE-pilot-based technique provides better MSE performance when compared to existing low-complexity high-PAPR pilot-based CFO (carrier frequency offset) estimator. Numerical study of the information rate performance of the TR-MRC receiver in imperfect CSI scenario with this CE-pilot based CFO estimator shows that it is more energy-and-spectrally efficient than existing low-complexity CFO estimator in massive MIMO systems. It is also observed that with this CE-pilot based CFO estimation, an $\mathcal{O}(\sqrt{M})$ array gain is achievable.
\end{abstract}

\vspace{-0.8 cm}
\begin{IEEEkeywords}

\vspace{-0.4 cm}
Spatially averaged, periodogram, low-complexity, carrier frequency offsets (CFOs), massive MIMO.
\end{IEEEkeywords}

\vspace{-0.5 cm}

%

\vspace{-0.4 cm}

\section{Introduction}
%
%
%
%
Massive multiple-input multiple-output (MIMO) system/large scale antennas system has recently been identified as one of the key technologies in the development of the next generation wireless communication network, because of its high energy and spectral efficiency \cite{Andrews}. Massive MIMO is a form of multi-user MIMO system, where the cellular base-station (BS) is equipped with a large array of antennas (of the order of hundreds), simultaneously serving several (of the order of tens) single antenna user terminals (UTs) in the same time-frequency resource \cite{Marzetta1}. Increasing number of BS antennas open up more available degrees of freedom, resulting in suppression of multi-user interference (MUI) and thus providing huge array gain. It has been shown that even in the imperfect CSI (channel state information) scenario, an $\mathcal{O}(\sqrt{M})$ array gain is achievable ($M$ is the number of BS antennas) \cite{Ngo1}. 

\par However all these results are based on coherent multi-user communication, for which perfect frequency synchronization is assumed. In practice, carrier frequency offsets (CFOs) between the received signal from UTs and the BS oscillator exist, which leads to degradation of system performance. Although various techniques have been developed over the past decade for conventional small scale MIMO systems \cite{Ghogho, Poor, Ma}, these techniques are not amenable to practical implementation in massive MIMO systems, due to prohibitive increase in their complexity with increasing number of UTs and also with increasing number of BS antennas.

\par In \cite{Larsson2}, the authors study CFO estimation in massive multi-user (MU) MIMO systems using an approximation to the joint maximum likelihood (ML) estimator. This technique requires multi-dimensional grid search and has exponential increase in complexity with increasing number of UTs. Recently in \cite{gcom2015} a low-complexity CFO estimation/compensation strategy has been suggested for massive MU-MIMO systems and impact of the residual CFO error on the information theoretic performance has been studied \cite{tvtrev2016}. However the CFO estimator discussed in \cite{gcom2015} requires high PAPR (peak-to-average-power ratio) pilots, which necessitates the use of linear power amplifiers (PAs), which are generally power inefficient. Since a massive MIMO system is expected to be highly energy efficient, it is desirable to use low PAPR pilots for CFO estimation as they allow the use of high efficiency non-linear PAs. In \cite{gcom2016}, a constant envelope (CE) pilot-based low-complexity CFO estimation algorithm has therefore been proposed and its mean squared error (MSE) performance has been studied. 

\par However it is not known whether coherent detection with this CE-pilot based CFO estimator can provide information rate performance (i.e. array gain and energy/spectral efficiency) similar to that with the high PAPR pilot-based CFO estimator in \cite{gcom2015, tvtrev2016}. These issues have been addressed in this paper. The major contributions are: (i) we study the complexity-performance (MSE performance) trade-off for this new CE-pilot based CFO estimator. Exhaustive numerical simulations show that for sufficiently large pilot length, the CE-pilot based CFO estimator has much better MSE performance than the high PAPR pilot-based CFO estimator presented in \cite{gcom2015}; (ii) we also study the impact of residual CFO errors on the information rate performance of the time-reversed maximum ratio combining (TR-MRC) receiver, with the CE-pilot based CFO estimator in the imperfect CSI scenario. Our study shows that an $\mathcal{O}(\sqrt{M})$ array gain is indeed achievable with this CE-pilot based CFO estimator, i.e., there is no degradation in the array gain performance when compared to the ideal/zero CFO scenario; (iii) finally from simulation studies it is observed that the CE-pilot based CFO estimator is more energy and spectrally efficient when compared to the CFO estimator in \cite{gcom2015}, specially when the coherence interval is sufficiently long. [\textbf{{Notations:}} $\E$ denotes the expectation operator and $(.)^{\ast}$ denotes the complex conjugate operator.]

 \begin{figure}[t]
 \centering
 \includegraphics[width= 4.5 in, height= 1.3 in]{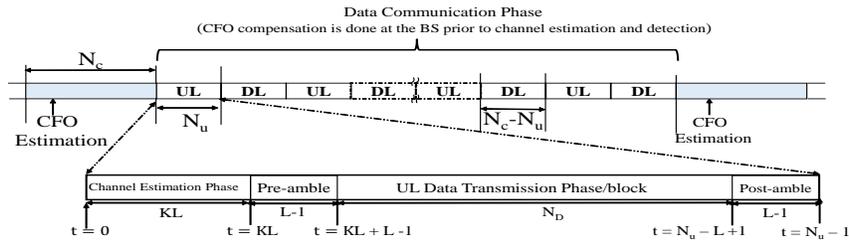}
 \caption {The communication strategy: CFO Estimation/Compensation and Data Communication. Here $N_c$ is the duration of coherence interval and the UL slot for data communication is $N_u$ channel uses.} 
 \label{fig:commstrat}
 \vspace{-0.5 cm}
 \end{figure}
 
 \vspace{-0.5 cm}

\section{System Model}

Let us consider a single-carrier single-cell massive MIMO BS, equipped with $M$ antennas, serving $K$ single antenna UTs simultaneously in the same time-frequency resource. Since a massive MIMO BS is expected to operate in time division duplexed (TDD) mode, each coherence interval is divided into an uplink (UL) slot, followed by a downlink (DL) slot. For coherent multi-user communication, frequency synchronization (i.e. CFO estimation/compensation) is important in massive MIMO systems. To this end, we consider a communication strategy, where the CFO estimation is performed in a special coherence slot (UL plus DL) before communication. In this slot, the UTs transmit special pilots to the BS. After CFO estimation, in the subsequent UL slots, at the BS, CFO compensation is performed, prior to channel estimation and UL receiver processing (see Fig.~\ref{fig:commstrat}). The special coherence slot for CFO estimation is repeated every few coherence intervals, depending on how fast the CFOs change.

\par The CFO estimation/compensation discussed in \cite{gcom2015} requires high PAPR pilots, which are susceptible to channel non-linearities. Since massive MIMO systems are highly energy efficient, it is desired that low-PAPR pilots be used for CFO estimation. In this paper, we consider constant envelope (CE) pilots. Specifically, for $K$ UTs, the $k^{\text{th}}$ UT would transmit a pilot $p_k[t] = e^{j\frac{2\pi}{K}(k - 1) t}$, where $k = 1, 2, \ldots, K$ and $t = 0, 1, \ldots, N-1$. Here $N \leq N_c$ is the pilot length and $N_c$ is the duration of a coherence interval. Assuming the channel to be frequency-selective with $L$ memory taps, the pilot signal received at time $t$ at the $m^{\text{th}}$ BS antenna would be given by

\vspace{-1 cm}

\begin{IEEEeqnarray}{rCl}
\label{eq:rxpilot}
r_m[t] & = & \sqrt{\pu}\sum\limits_{q = 1}^{K}\sum\limits_{l = 0}^{L-1}h_{mq}[l]\, e^{j[\frac{2\pi}{K}(q - 1)(t - l) + \omega_q t]} \, + \, n_m[t] =  \sqrt{\pu} \sum\limits_{q=1}^{K} H_{mq} \, e^{j[\frac{2\pi}{K}(q - 1) + \omega_q]t} \, + \, n_m[t],
\IEEEeqnarraynumspace
\end{IEEEeqnarray}

\vspace{-0.2 cm}

\noindent where $H_{mq} \Define \sum\limits_{l = 0}^{L-1} h_{mq}[l] \, e^{-j \frac{2\pi}{K}(q-1)l}$ and $\omega_q$ is the CFO of the $q^{\text{th}}$ UT. Here $\pu$ is the average power transmitted by each UT and $h_{mk}[l] \sim \mathcal{C}\mathcal{N}(0, \sigma_{hkl}^2)$ is the independent channel gain coefficient from the single-antenna of the $k$-th UT to the $m$-th antenna of the BS at the $l$-th channel tap. Also, $\{\sigma_{hkl} > 0\}, \, (l = 0, 1, \ldots, L-1; \, k = 1, 2, \ldots, K)$ is perfectly known at the BS and models the power delay profile (PDP) of the channel. 

\vspace{-0.3 cm}

\subsection{Low-Complexity CFO Estimation Using Spatially Averaged Periodogram}

From \eqref{eq:rxpilot} it is clear that the signal received at the BS is simply a sum of complex sinusoids with additive noise. Specifically, the frequency of the sinusoid received from the $k^{\text{th}}$ UT is $\frac{2\pi}{K}(k - 1) + \omega_k$. Intuitively an estimate of the CFO of the $k^{\text{th}}$ UT would be the difference between the frequency of the transmitted pilot (i.e. $\frac{2\pi}{K}(k - 1)$) and the estimated frequency of the sinusoid received at the BS from the $k^{\text{th}}$ user. An attractive low-complexity alternative to the high-complexity joint ML frequency estimator is the periodogram technique \cite{Stoica}, which simply requires to compute periodogram of the received signal and choose the $K$ largest peaks as the estimates of the $K$ frequencies. In massive MIMO systems, the received signal power at each BS antenna is expected to be small and therefore we propose to perform spatial averaging of the periodogram, computed separately at each of the $M$ BS antennas.

\par Assuming the CFOs from all UTs lie within the range $[-\Delta_{\text{max}}, \Delta_{\text{max}}]$ (where $\Delta_{\text{max}}$ is the maximum CFO for any UT), the frequency of the received pilot from the $k^{\text{th}}$ UT would lie in the interval $[\frac{2\pi}{K}(k - 1) - \Delta_{\text{max}}, \frac{2\pi}{K}(k - 1) + \Delta_{\text{max}}]$. Since $\Delta_{\text{max}} \ll \frac{\pi}{K}$ in practice\footnote[1]{For a massive MIMO system with carrier frequency $f_c = 2$ GHz, communication bandwidth $B_w = 1$ MHz and maximum frequency offset $\kappa = 0.1$ PPM of $f_c$ \cite{Weiss}, the maximum CFO is given by $\Delta_{\text{max}} = 2\pi \kappa f_c/ B_w = \frac{\pi}{2500}\ll \frac{\pi}{K}$, where in massive MIMO systems, $K$ is only of the order of tens. A detailed discussion on the range and values of CFOs in massive MIMO systems is given in \cite{gcom2015}.}, these intervals for different UTs would be non-overlapping. Therefore instead of computing the periodogram over the entire interval $[-\pi, \pi]$, we only compute the periodogram in the interval $[\frac{2\pi}{K}(k - 1) - \Delta_{\text{max}}, \frac{2\pi}{K}(k - 1) + \Delta_{\text{max}}]$ over a fine grid (i.e. at discrete frequencies). Thus the proposed CFO estimator for the $k^{\text{th}}$ UT is given by

\vspace{-0.9 cm}

\begin{IEEEeqnarray}{rCl}
\label{eq:periodogram}
\widehat{\omega}_k = \arg \max\limits_{\varOmega(i) \in \varXi} \, \underbrace{\overbrace{\frac{1}{M}\sum\limits_{m = 1}^{M}}^{\substack{\text{Spatial}\\ \text{averaging}}}\overbrace{\frac{1}{N}\Big | \sum\limits_{t = 0}^{N-1}r_m[t] \, e^{-j[\frac{2\pi}{K}(k - 1) + \varOmega(i)]t}\Big |^2}^{\text{Periodogram computed at the $m^{\text{th}}$ BS antenna}}}_{ \Define \, \Phi_{k}(\varOmega(i))},
\IEEEeqnarraynumspace
\end{IEEEeqnarray}

\vspace{-0.3 cm}

\noindent where $\varXi \Define \{\varOmega(i) \Define \frac{2\pi}{N^{\alpha}}i \Big | |i| \leq T_0\}$, $T_0 \Define \lceil \frac{\Delta_{\text{max}}}{2\pi} N^{\alpha}\rceil$ and $\varOmega(i)$ denotes the discrete frequencies where the periodogram is computed. Note that the parameter $\alpha$ serves as the level of resolution of discrete frequencies in the set $\varXi$. Clearly with increasing $\alpha$ for a fixed $N$, the resolution of the CFO estimator would increase and therefore the MSE of CFO estimation, $\epsilon \Define \E[(\widehat{\omega}_k - \omega_k)^2]$ would decrease.

\vspace{-0.5 cm}

\subsection{Performance-Complexity Trade-off}

From \eqref{eq:periodogram} it is clear that the total number of operations required to compute $\Phi_k(\varOmega(i))$ is $\mathcal{O}(MN)$ where $|i| \leq T_0$. Clearly the total number of operations per-channel use for all $K$ UTs would be $\mathcal{O}(MKT_0)$, where $T_0 = \lceil \frac{\Delta_{\text{max}}}{2\pi}N^{\alpha} \rceil$. Clearly with $M$, $K$ and $N$ fixed and increasing $\alpha$, the complexity of the CFO estimator increases, while the MSE of CFO estimation decreases. It is however observed that with increasing $\alpha$, the incremental reduction in the MSE of CFO estimation becomes negligible, when $\alpha$ is sufficiently large. In this paper, for a given pilot length $N$, we therefore choose $\alpha$ to be the smallest value such that $|(\epsilon(\alpha) - \epsilon(\alpha + \Delta \alpha))/ \epsilon(\alpha)| < \delta$, for a given $\delta > 0$ and $\Delta \alpha > 0$. Here, $\epsilon(\alpha)$ is the MSE of CFO estimation for a given $\alpha$.

\begin{figure}[t]
\centering
\includegraphics[width= 4.5 in, height= 2.6 in]{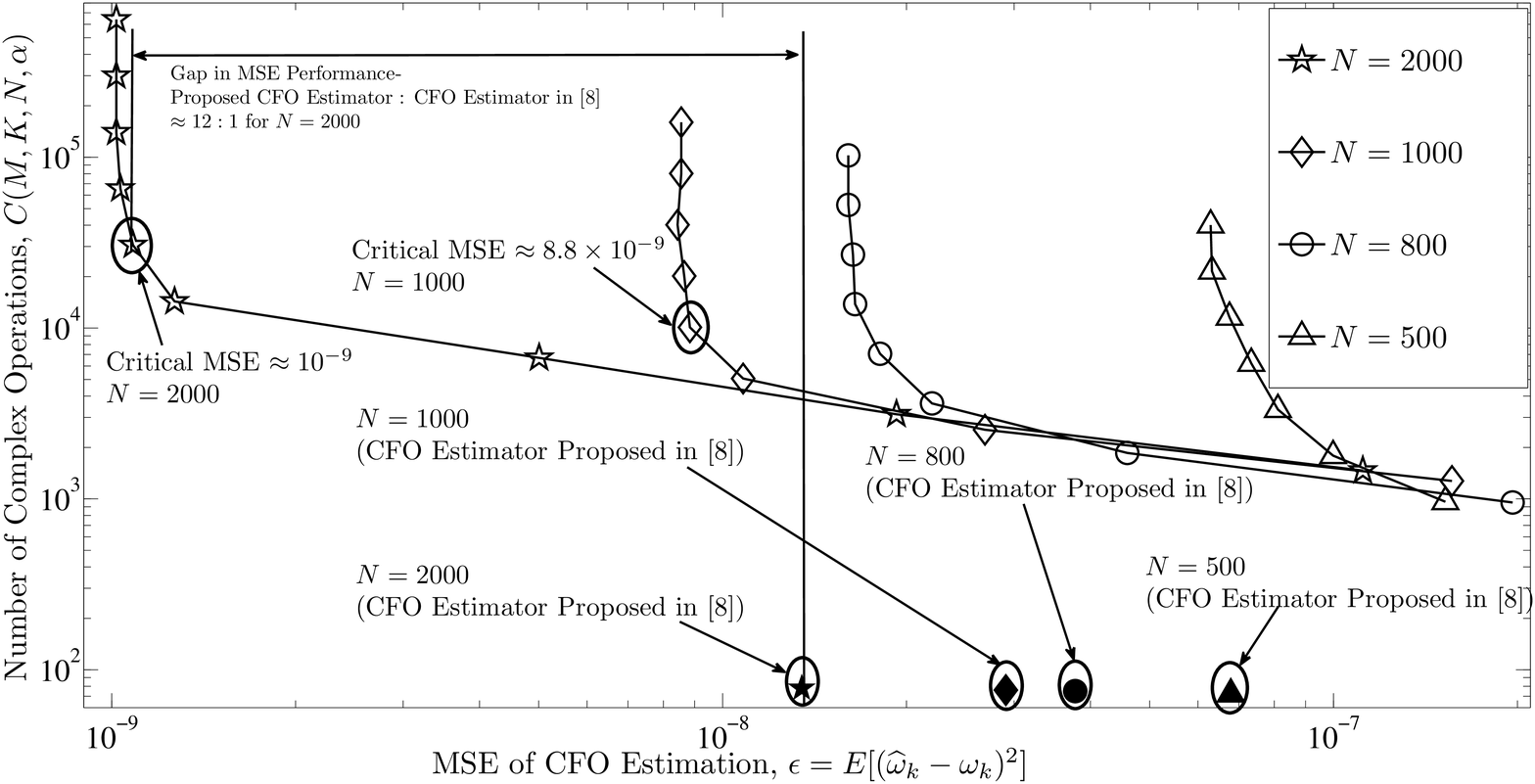}
\caption {Plot of the variation in the total number of complex operations $C(M, K, N, \alpha)$ with decreasing MSE of CFO estimation, for fixed $M = 80$, $K = 10$, $L = 5$, SNR = $\gamma = \frac{\pu}{\sigma^2} = -10$ dB and $N = 500$, $800$, $1000$ and $2000$ respectively.} 
\label{fig:compvsres}
\vspace{-0.5 cm}
\end{figure}

\par In Fig.~\ref{fig:compvsres} we plot the variation in the complexity (i.e. the total number of complex floating point operations, $C(M, K, N, \alpha)$) with decreasing MSE of CFO estimation for a fixed $M = 80$, $K = 10$, $L = 5$, SNR $\gamma \Define \frac{\pu}{\sigma^2} = -10$ dB and $N = 500$, $800$, $1000$ and $2000$. Note that for a fixed $N$, with decreasing MSE, $C(M, K, N, \alpha)$ increases and below a critical value of the MSE, the change in MSE becomes negligible with increasing $C(M, K, N, \alpha)$. Clearly this critical value of MSE is a good operating point in terms of the complexity-performance trade-off. We also plot the required number of complex operations for the high PAPR pilot-based CFO estimator proposed in \cite{gcom2015}. Note that when $N$ is small, the high PAPR pilot-based CFO estimator discussed in \cite{gcom2015} is better both in terms of complexity and MSE performance. However with increasing $N$, it is observed that the proposed CE-pilot based CFO estimator quickly out-performs the high PAPR pilot-based CFO estimator discussed in \cite{gcom2015} in terms of MSE performance (see the MSE performance gap for $N = 2000$ in Fig.~\ref{fig:compvsres}).

\vspace{-0.5 cm}

\section{Information Rate Analysis}

After the CFO estimation phase, the conventional data communication starts at $t = 0$ of the next UL slot (see Fig.~\ref{fig:commstrat}). The UTs transmit pilots for channel estimation sequentially in time. Specifically, the $k^{\text{th}}$ UT transmits an impulse of amplitude $\sqrt{KL \pu}$ at time $t = (k - 1)L$ and zeros elsewhere. Therefore the received pilot at the $m^{\text{th}}$ BS antenna at time $t = (k - 1)L +l$ is given by $r_m[(k - 1)L + l] = \sqrt{KL \pu}\, h_{mk}[l] \, e^{j \omega_k[(k - 1)L + l]} + n_m[(k - 1)L + l]$, where $m = 1, 2, \ldots, M$, $l = 0, 1, \ldots, L-1$ and $k = 1, 2, \ldots, K$. To estimate the channel gain coefficient, we first perform CFO compensation for the $k^{\text{th}}$ UT by multiplying $r_m[(k - 1)L + l]$ with $e^{-j\widehat{\omega}_k[(k - 1)L +l]}$ and then computing the channel estimate as $\widehat{h}_{mk}[l] \Define r_m[(k - 1)L +l]e^{-j\widehat{\omega}_k[(k-1)L + l]}/\sqrt{KL \pu} = \widetilde{h}_{mk}[l] + \frac{1}{KL \pu}\widetilde{n}_m[(k-1)L + l]$. Here $\widetilde{n}_m[(k-1)L+l] \Define n_m[(k-1)L+l]e^{-j\widehat{\omega}_k[(k-1)L+l]} \sim \mathcal{C}\mathcal{N}(0, \sigma^2)$ and $\widetilde{h}_{mk}[l] \Define h_{mk}[l]e^{-j\Delta \omega_k [(k-1)L+l]} \sim \mathcal{C}\mathcal{N}(0,\sigma_{hkl}^2)$ is the effective channel gain coefficient and $\Delta \omega \Define \widehat{\omega}_k - \omega_k$ is the residual CFO after compensation.\footnote[2]{Both $h_{mk}[l]$ and $n_m[(k-1)L+l]$ have uniform phase distribution (i.e. circular symmetric) and are independent of each other. Clearly, rotating these random variables by fixed angles (for a given realization of CFOs and its estimates) would not change the distribution of their phases and they will remain independent. Therefore the distribution of $\widetilde{h}_{mk}[l]$ and $\widetilde{n}_{mk}[(k-1)L+l]$ would be same as that of $h_{mk}[l]$ and $n_m[(k-1)L+l]$ respectively.}

\par After channel estimation ($KL$ channel uses) and ($L - 1$) channel uses of preamble transmission\footnote[3]{The symbols transmitted in the pre-amble and post-amble sequences are independent and identically distributed (i.i.d.) with the same distribution as that of the information symbols and average power $\pu$. This is required to guarantee the correctness of the achievable information rate expression.}, the UL data transmission block of $N_D$ channel uses starts at $t = KL + L -1$ (see Fig.~\ref{fig:commstrat}). Let $x_k[t] \sim \mathcal{C}\mathcal{N}(0,1)$ be the i.i.d. information symbol transmitted by the $k^{\text{th}}$ UT at the $t^{\text{th}}$ channel use and $\pu$ be the average power transmitted by each UT. Therefore the received signal at the $m^{\text{th}}$ BS antenna at time $t$ is given by $r_m[t] = \sqrt{\pu}\sum_{q = 1}^{K} \sum_{l = 0}^{L - 1}h_{mq}[l]x_q[t - l]e^{j \omega_q t} + n_m[t]$, where $t = KL+L-1, \ldots, (N_u-L)$ and $N_u = KL + N_D + 2(L-1)$ is the duration of the UL slot. To detect $x_k[t]$, we first perform CFO compensation for the $k^{\text{th}}$ UT, followed by time reversed maximum ratio combining (TR-MRC) \cite{Phasenoise}. Therefore the output of the TR-MRC receiver for the $k^{\text{th}}$ UT at time $t$ is given by 

\vspace{-0.9 cm}

{\begin{IEEEeqnarray}{rCl}
\label{eq:xkhat}
\nonumber \widehat{x}_k[t] & \Define & \sum\limits_{m = 1}^{M}\sum\limits_{l = 0}^{L - 1}\widehat{h}_{mk}^{\ast}[l] \underbrace{ r_m[t+l]e^{-j\widehat{\omega}_k(t+l)}}_{\text{CFO compensation}} = \underbrace{\sqrt{\pu}\E \Big[\sum\limits_{m = 1}^{M} \sum\limits_{l = 0}^{L-1}|\widetilde{h}_{mk}[l]|^2 \, e^{-j\Delta\omega_k(t - (k-1)L)}\Big]x_k[t] }_{\Define \, \text{ES}_k[t]} + \text{IN}_k[t]\\
 &  & \, + \,\, \underbrace{\sqrt{\pu} \sum\limits_{m = 1}^{M} \sum\limits_{l = 0}^{L-1}|\widetilde{h}_{mk}[l]|^2 \, e^{-j\Delta\omega_k(t - (k-1)L)} x_k[t] - \text{ES}_k[t]}_{\Define \text{SIF}_k[t]}\,\,\, =  \text{ES}_k[t] + \text{EW}_k[t],
\end{IEEEeqnarray}}

\vspace{-0.6 cm}

\noindent where $\text{ES}_k[t]$ is the effective information signal and $\text{IN}_k[t]$ denotes the sum of inter-symbol interference (ISI), multi-user interference (MUI) and AWGN components of the received signal. From numerical simulations, it is observed that the statistics of $\text{EW}_k[t] \Define \text{SIF}_k[t] + \text{IN}_k[t]$ varies with $t$. However, for a given $t$, the realization of $\text{EW}_k[t]$ is i.i.d. across multiple UL data transmission blocks. Therefore when viewed across multiple coherence blocks, for each channel use in \eqref{eq:xkhat}, we essentially have a SISO (single-input single-output) channel.\footnote[4]{We therefore have separate codebooks, one for each channel use. This coding strategy has also been used in \cite{Phasenoise,tvtrev2016}. Note that in practice, since the statistics of $\text{EW}_k[t]$ varies slowly with $t$, in addition to coding across different UL data transmission blocks, one can also code across consecutive channel uses in the same UL data transmission block.} From exhaustive numerical simulations it follows that $\E[\text{ES}_k[t] \text{EW}_k^{\ast}[t]] = 0$, i.e., the overall noise and interference is uncorrelated with the effective information signal. With i.i.d. Gaussian information symbols, we can obtain a lower bound on the achievable information rate of the effective channel in \eqref{eq:xkhat}, by considering the worst case uncorrelated additive noise (in terms of mutual information), which would have the same variance as $\text{EW}_k[t]$ \cite{Hasibi2}. Thus an achievable information rate for the $k^{\text{th}}$ UT is given by $I_k = \frac{1}{N_u}\sum_{t = KL+L-1}^{N_u - L}\log_2(1 + \text{SINR}_k[t])$, where $\text{SINR}_k[t] \Define {\E[|\text{ES}_k[t]|^2]}/{\E[|\text{EW}_k[t]|^2]}$.

\begin{remark}
\label{arraygain}
(Achievable Array Gain)
\normalfont
From exhaustive numerical simulations it can be shown that the ISI, MUI and AWGN components in $\text{IN}_k[t]$ are uncorrelated since $x_k[t]$ are all i.i.d. Also, the variances of these components do not depend on the residual CFOs. Further from numerical simulations, we observe that the variance of ISI and MUI components would vanish with increasing number of BS antennas $M \to \infty$ and the transmit SNR $\gamma$ decreasing as $\frac{1}{\sqrt{M}}$, while the variance of AWGN component in $\text{IN}_k[t]$ approaches a constant value \cite{tvtrev2016, Marzetta1} (fixed $K$, $L$ and $N$). Therefore the residual CFO error can impact the SINR only through the variances of $\text{ES}_k[t]$ and $\text{SIF}_k[t]$. Since the MSE of CFO estimation converges to a constant value with increasing $M \to \infty$ and $\gamma \propto \frac{1}{\sqrt{M}}$ (see Remark~3 in \cite{gcom2016}), we can conclude that with $\gamma \propto \frac{1}{\sqrt{M}}$, the overall variance of $\text{ES}_k[t]$ and $\text{SIF}_k[t]$ would also converge to a constant value with increasing $M \to \infty$. Thus for a fixed $N$, $K$ and $L$, the achievable information rate would approach a constant value with decreasing transmit SNR $\gamma \propto \frac{1}{\sqrt{M}}$ and increasing $M \to \infty$. This shows that an $\mathcal{O}(\sqrt{M})$ array gain is also achievable with the new CFO estimation/compensation technique proposed in this paper. This conclusion is also supported through Table~\ref{table:snrM}. \hfill \qed
\end{remark}

\vspace{-0.6 cm}

\section{Numerical Results and Discussions}

In this section, we compare the performance of the TR-MRC receiver with the proposed CFO estimator (i.e. CE-pilot based spatially averaged periodogram) to the performance of (i) the TR-MRC receiver with the high PAPR pilot-based CFO estimator in \cite{gcom2015}, and (ii) the TR-MRC receiver in an ideal/zero CFO scenario. For Monte-Carlo simulations, we assume the following: carrier frequency $f_c = 2$ GHz, communication bandwidth $B_w = 1$ MHz, maximum CFO $= 0.1$ PPM of $f_c$, i.e., $\Delta_{\text{max}} = \frac{\pi}{2500}$, pilot length $N = 2000$ and maximum delay spread $T_d = 5 \mu$s. Clearly, $L = T_d B_w = 5$. At the start of every CFO estimation phase, the CFOs $\omega_k$ ($k = 1, 2, \ldots, K$) assume new values (independent of the previous ones), uniformly distributed in $[-\frac{\pi}{2500}, \frac{\pi}{2500}]$. Further the PDPs are assumed to be same for all UTs and is given by $\sigma_{hkl}^2 = 1/L$, where $l = 0, 1, \ldots, L-1$.

\begin{figure}[t]
\centering
\includegraphics[width= 4.5 in, height= 2.4 in]{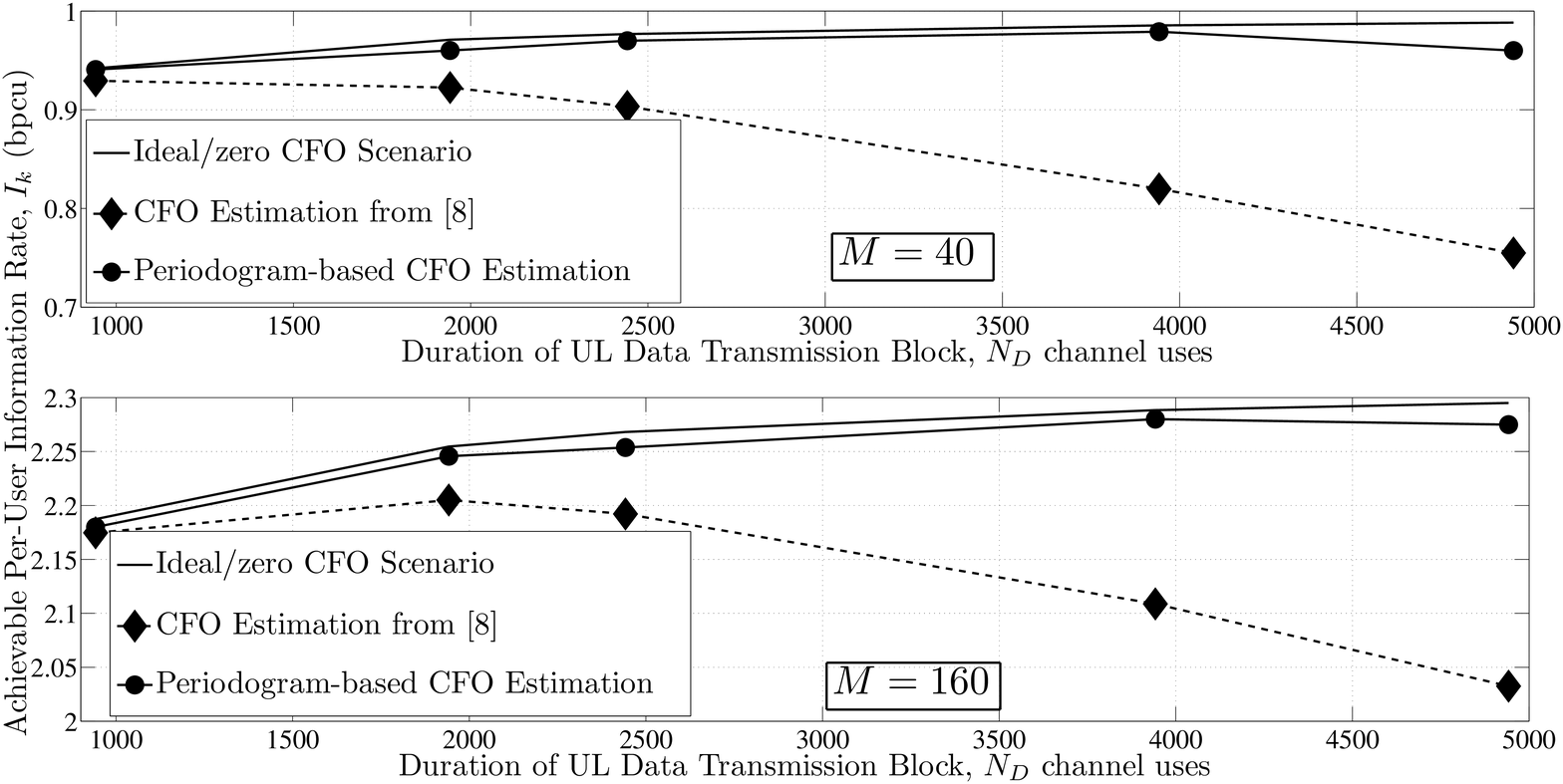}
\caption {Plot of the variation of achievable per-user information rate versus duration of UL data transmission block, $N_D$ channel uses with fixed $N = 2000$, $K = 10$, $L = 5$, $SNR = -10$ dB and $M = 40, 160$.} 
\label{fig:infoblock}
\vspace{-0.5 cm}
\end{figure}

\begin{savenotes}
\begin{table}[b]
\caption[position=top]{{\textsc{Minimum Required SNR $\gamma = \frac{\pu}{\sigma^2}$ to achieve a fixed Per-User Information rate $I_k = 1$ bpcu, $K = 10$, $N = 2000$, $L = 5$ and $N_c = 10000$.}}}
\label{table:snrM}
\centering
\begin{tabular}{| c | c | c | c | c | c |}
\hline
M & 40 & 80 & 160 & 320 & 640\\
\hline
\text{SNR} & -9.9 & -12.53 & -14.7 & -16.6 & -18.38 \\                 
\hline
\end{tabular}
\end{table}
\end{savenotes}

\par In Fig.~\ref{fig:infoblock} we plot the variation in the achievable per-user information rate with increasing duration of the UL data transmission block ($N_D$) for $M = 40, 160$. It is observed that with increasing $N_D$, the achievable information rate initially increases, but later starts to decrease. This is due to the fact that with increasing $N_D$, the channel estimates used for coherent detection at the BS becomes stale (i.e. due to the CFO, the phase error between the acquired channel estimates and the channel gain in the received information signal increases with increasing time lag between the channel estimation phase and the time instance when the information symbol is received). Note that with the CE pilot-based CFO estimator the achievable information rate performance is very close to the ideal/zero CFO scenario (even when $N_D$ is very high), while the performance with the high PAPR pilot-based CFO estimator presented in \cite{gcom2015} degrades rapidly. Equivalently, this also shows that the CE-pilot based CFO estimator is more energy efficient than the high PAPR pilot-based CFO estimator presented in \cite{gcom2015}. Also in Table~\ref{table:snrM}, for the CE-pilot based CFO estimator, we show the variation in the minimum required SNR with increasing $M$ for a fixed desired per-user information rate of $1$ bpcu (bits per channel use). Note that with $M \to \infty$, the required SNR decreases by almost $1.5$ dB with every doubling in $M$ (see the variation in SNR for $M = 320$ and $M = 640$). This supports our conclusion in Remark~\ref{arraygain}.





%

%


\ifCLASSOPTIONcaptionsoff
  \newpage
\fi



%

        \vspace{-0.7 cm}

\bibliographystyle{IEEEtran}
\bibliography{IEEEabrvn,mybibn}

\nocite{gcom2016,Weiss,Hasibi2}

\end{document}